\documentclass[11pt]{article}
\begin{document}
\vspace*{2cm}

\begin{center}
{NAMBU-GOLDSTONE MODES IN GRAVITATIONAL THEORIES\\
WITH SPONTANEOUS LORENTZ BREAKING}
\footnote{Presented at the meeting {\it From Quantum to Cosmos},
Washington, D.C., May 2006.}
\end{center}

\begin{center}
{R.\ BLUHM}
\end{center}

\begin{center}
{Physics Department, Colby College,
Waterville, ME 04901, USA}
\end{center}

\begin{abstract}
\noindent
Spontaneous breaking of Lorentz symmetry has been suggested as a 
possible mechanism that might occur in the context of a fundamental 
Planck-scale theory, such as string theory or a quantum theory of gravity.  
However, if Lorentz symmetry is spontaneously broken, two sets of
questions immediately arise: what is the fate of the Nambu-Goldstone 
(NG) modes, and can a Higgs mechanism occur?
A brief summary of some recent work looking at these questions
is presented here.

\end{abstract}

\section{Introduction}	

In gauge theory, spontaneous symmetry breaking has well known consequences.  
The Goldstone theorem states that when a continuous global symmetry is
spontaneously broken, massless Nambu-Goldstone (NG) modes appear.\cite{ng}
On the other hand, if the symmetry is local, 
then a Higgs mechanism can occur in which the gauge
fields acquire mass.\cite{hm}

In this work,
these processes are examined for the case where the symmmetry
is Lorentz symmetry.
In flat spacetime, Lorentz symmetry is a global symmetry.
Therefore, if the symmetry is spontaneously broken,
it is expected that massless NG modes should appear. 
However, in curved spacetime, in a gravitational theory, 
Lorentz symmetry is a local symmetry. 
It is in this context that the possibility of a Higgs mechanism arises.

The question of what the fate of the NG modes is when Lorentz symmetry
is spontaneously broken was recently examined,\cite{rbak05}
including the possibility of a Higgs mechanism.
It is mainly the results of this work that are summarized here.
However,
the original motivation for considering the possibility that Lorentz
symmetry might be spontaneously broken stems from the work 
in the late 1980s of Kosteleck\'y and Samuel.\cite{ks}
For example, they found that mechanisms occurring in the context of string field 
theory can lead to spontaneous Lorentz breaking.
This led them to propose a vector model with spontaneous Lorentz 
violation, now known as a bumblebee model, which can be used to
study the gravitational implications of spontaneous Lorentz violation.
This model is summarized here as well, as are some of the original results
of Kosteleck\'y and Samuel concerning an alternative Higgs mechanism 
involving the metric.
A number of additional studies concerning the bumblebee model have 
been carried out in recent years,\cite{bb1,bb2}
with many of them leading to new ideas concerning modified gravity
and new phenomenological tests of relativity theory.
However, space limitations do not permit a full review.
Instead, the focus here will be on the NG modes and gravitational Higgs mechanism.

\section{Spontaneous Lorentz Breaking}

Lorentz symmetry is spontaneously broken when a local tensor 
field acquires a vacuum expectation value (vev),
\begin{equation}
<T_{abc}> \, = t_{abc}
\quad .
\label{Tvev}
\end{equation}       
The vacuum of the theory then has preferred spacetime directions, 
which spontaneously breaks the symmetry.

In curved spacetime, the Lorentz group acts locally at each spacetime point.  
In addition to being locally Lorentz invariant, a gravitational theory is also
invariant under diffeomorphisms.  
There are therefore two relevant symmetries, 
and it is important to consider them both.
While Lorentz symmetry acts in local frames, and transforms 
tensor components with respect to a local basis, e.g., $T_{abc}$
(where Latin indices denote components with respect to a local frame), 
diffeomorphism act in
the spacetime manifold and transform components $T_{\lambda\mu\nu}$ 
defined (using Greek indices) with respect to the spacetime coordinate system.
These local and spacetime tensor components are linked by a vierbein.
For example,
the spacetime metric and local Minkowski metric are related by
\begin{equation}
g_{\mu\nu} = e_\mu^{\,\,\, a} e_\nu^{\,\,\, b} \eta_{ab} 
\quad .
\label{vier}
\end{equation}       
In a similar way,
spacetime tensor components are related to the components in a
local frame using the vierbein,
\begin{equation}
T_{\lambda\mu\nu} \, = e_\lambda^{\,\,\, a} e_\mu^{\,\,\, b} e_\nu^{\,\,\, c} \, t_{abc}
\quad .
\label{T}
\end{equation}       

There are a number of reasons why it is natural to use a vierbein formalism 
to consider local Lorentz symmetry in the context of a gravitational theory.  
First, the introduction of vierbeins allows spinors to be incorporated into the theory.  
The vierbein formalism also naturally parallels gauge theory,
with Lorentz symmetry acting as a local symmetry group.
The spin connection $\omega_\mu^{\,\,\, ab}$ enters in covariant
derivatives that act on local tensor components and plays the role of
the gauge field for the Lorentz symmetry.
In contrast,
the metric excitations act as the gauge
fields for the diffeomorphism symmetry.
When working with a vierbein formalism,
there are primarily two distinct geometries that must be distinguished.
In a Riemannian geometry (with no torsion), 
the spin connection is nondynamical.  
It is purely an auxiliary field that does not propagate.  
However, in a Riemann-Cartan geometry (with nonzero torsion), 
the spin connection must be treated as independent degrees of freedom
that in principle can propagate.\cite{rcst}

\section{Nambu-Goldstone Modes}

Consider a theory with a tensor vev in a local Lorentz frame,
$<T_{abc}> \, = t_{abc}$,       
which spontaneously breaks Lorentz symmetry.  
The vacuum value of the vierbein is also a constant or fixed function, 
e.g., for simplicity
consider a background Minkowski spacetime where
\begin{equation}
<e_\mu^{\,\,\, a}> \, = \delta_\mu^{\,\,\, a}
\quad .
\label{evev}
\end{equation}       
The spacetime tensor therefore has a vev as well,
\begin{equation}
<T_{\lambda\mu\nu}> \, = t_{\lambda\mu\nu}
\quad .
\label{Tmunuvev}
\end{equation}     
This means that diffeomorphisms are also spontaneously broken.
Our first result is therefore that
{\it spontaneous breaking of local Lorentz symmetry
implies spontaneous breaking of diffeomorphisms.}

The spontaneous breaking of these symmetries implies that NG
modes should appear (in the absence of a Higgs mechanism).
This immediately raises the question of how many NG modes appear. 
In general, there can be up to as many NG modes as there are broken symmetries. 
Since the maximal symmetry-breaking case would yield six broken Lorentz generators 
and four broken diffeomorphisms,
there can therefore be up to ten NG modes.
A natural follow-up question is to ask where these modes reside.
In general, this depends on the choices of gauge.  
However, one natural choice is to put all the NG modes into the vierbein,
as a simple counting argument shows is possible.
The vierbein $e_\mu^{\,\,\, a}$ has 16 components.  
With no spontaneous Lorentz violation, 
the six Lorentz and four diffeomorphism degrees
of freedom can be used to reduce the vierbein down to six independent degrees
of freedom.  
(Note that a general gravitational theory can have six propagating metric modes; 
however, general relativity is special in that there are only two). 
In contrast, in a theory with spontaneous
Lorentz breaking, 
where all ten spacetime symmetries have been broken,
up to all ten NG modes can potentially propagate.
Thus, our second result is that
{\it in a theory with spontaneous Lorentz breaking,
up to ten NG modes can appear and all of them can naturally
be incorporated as degrees of freedom in the vierbein.}

\section{Bumblebee Model}

The simplest case of a theory with spontaneous Lorentz breaking is 
a bumblebee model.\cite{ks}  
These are defined as theories in which a vector field $B_\mu$ acquires a vev,
\begin{equation}
<B_\mu> \, = b_\mu
\quad .
\label{Bvev}
\end{equation}       
The vev can be induced by a potential $V$ in the Lagrangian that has a minimum for 
nonzero values of the vector field.  
A simple example of a bumblebee model has the form
${\cal L} = {\cal L}_{\rm G} + {\cal L}_{\rm B} + {\cal L}_{\rm M}$,
where ${\cal L}_{\rm G}$
describes the pure-gravity sector,
${\cal L}_{\rm M}$ describes the matter sector,
and (choosing a Maxwell form for the kinetic term)
\begin{equation}
{\cal L}_{\rm B} = \sqrt{-g} \left( - \frac 1 4 B_{\mu\nu} B^{\mu\nu} 
- V(B_\mu) + B_\mu J^\mu \right) ,
\label{BBL}
\end{equation}
describes the bumblebee field.
Here,
$J^\mu$ is a matter current,
and the bumblebee field strength is
$B_{\mu\nu} = D_\mu B_\nu - D_\nu B_\mu$,
which in a Riemann spacetime (with no torsion) reduces to
$B_{\mu\nu} = \partial_\mu B_\nu - \partial_\nu B_\mu$.
(For simplicity, we are neglecting additional possible interactions between the
curvature tensor and $B_\mu$).

The potential $V$ depends on $B_\mu$ and the metric $g_{\mu\nu}$.
It is chosen so that its minimum occurs when $B_\mu$ and $g_{\mu\nu}$
acquire nonzero vacuum expectation values.
For a general class of theories,
$V$ is a function of $(B_\mu B^\nu \pm b^2)$,
with $b^2 > 0$ equaling a constant,
and where the minimum of the potential occurs when 
$B_\mu g^{\mu\nu} B_\nu \pm b^2 = 0$.
The vacuum solutions for $B_\mu$
(which can be timelike or spacelike depending on the choice of sign)
as well as for $g_{\mu\nu}$
must be nonzero and therefore spontaneously break both Lorentz and
diffeomorphism symmetry.

Among the possible choices for the potential are a sigma-model potential
$V = \lambda (B_\mu B^\nu \pm b^2)$, 
where $\lambda$ is a Lagrange-multiplier field,
and a squared potential
$V = \frac 1 2 \kappa (B_\mu B^\nu \pm b^2)^2$,
where $\kappa$ is a constant (of mass dimension zero).
In the former case,
only excitations that stay within the potential minimum
(the NG modes) are allowed by the constraint imposed by $\lambda$.
However,
in the latter case,
excitations out of the potential minimum are possible as well.
In either of these models, 
three Lorentz symmetries and one diffeomorphism are broken, 
and therefore up to four NG modes can appear.  
However, the diffeomorphism NG mode does not propagate.\cite{rbak05}
It drops out of all of the kinetic terms and is purely an auxiliary field.  
In contrast, the Lorentz NG modes do propagate.  
They comprise a massless vector,
with two independent transverse degrees of freedom (or polarizations).
Indeed,
they are found to propagate just like a photon.

\section{Photons and Lorentz Violation}

We find that the NG modes resulting from spontaneous local Lorentz violation 
can lead to an alternative explanation for the existence of massless photons
(besides that of U(1) gauge invariance).
Prevoius links between QED gauge fields, fermion composites, and the NG modes 
had been uncovered in flat spacetime (with global Lorentz symmetry).\cite{ngphoton}  
Here, we propose a theory with just a vector field (but no U(1) gauge symmetry)
giving rise to photons in the context of a gravitational theory
where local Lorentz symmetry is spontaneously broken.\cite{rbak05}
Defining $B_\mu - b_\mu = A_\mu$, 
we find at lowest order that the Lorentz NG excitations propagate 
as transverse massless modes obeying an axial gauge condition,  
$b_\mu A^\mu = 0$. 
Hence, in summary, our third result is that
{\it spontaneous local Lorentz violation may provide an alternative explanation for
massless photons.}
In the bumblebee model,
the photon fields
couple to the current $J_\mu$ as conventional photons, 
but also have additional Lorentz-violating background interactions 
like those appearing in the Standard-Model Extension(SME).\cite{sme}
By studying these additional interactions,
signatures can be searched for that might ultimately
distinguish between a photon theory based on local Lorentz breaking
from that of conventional Einstein-Maxwell theory.

\section{Higgs Mechanisms}

Since there are two sets of broken symmetries (Lorentz and diffeomorphisms) 
there are potentially two associated  Higgs mechanisms. 
However,
in addition to the usual Higgs mechanism
(in which a gauge-covariant-derivative term generates a mass term
in the Lagrangian),
it was shown\cite{ks}
that an alternative Higgs mechanism can occur due to the
gravitational couplings that appear in the potential $V$.

First,
consider the case of diffeomorphisms.
Here,
it was shown 
that the usual Higgs mechanism involving the metric does not occur.\cite{ks}  
This is because the mass term that is generated by covariant derivatives
involves the connection,  which consists of derivatives of the metric
and not the metric itself.
As a result,
no  mass term for the metric is generated following the usual Higgs prescription.
However, it was also shown that because of the form of the potential, 
e.g., $V=V(B_\mu g^{\mu\nu} B_\nu + b^2)$, 
quadratic terms for the metric can arise, 
resulting in an alternative form of the Higgs mechanism.\cite{ks}
These can lead to mass terms that can potentially modify gravity
in a way that avoids the van Dam, Veltmann, and Zakharov discontinuity.\cite{vdvz}
Summarizing the case of diffeomorphisms,
we have that 
{\it there is no conventional Higgs mechanism for the graviton;  
however, mass terms for the metric may arise from the potential 
$V$ in an alternative mechanism.}

In contrast, for the case of Lorentz symmetry, 
it is found that a conventional Higgs mechanism can occur.\cite{rbak05}
In this case the relevant gauge field (for the Lorentz symmetry) 
is the spin connection.  
This field appears directly in covariant derivatives acting on local tensor
components,
and for the case where the local tensors acquire a vev,
quadratic mass terms for the spin connection can be generated
following a similar prescription as in the usual Higgs mechanism.
For example. in the bumblebee model, 
using a unitary gauge, 
the kinetic terms involving $B_{\mu\nu}$ generate quadratic mass terms
for the spin connection $\omega_\mu^{\,\,\, ab}$. 
However, a viable Higgs mechanism involving the spin connection can 
only occur if the spin connection is a dynamical field.  
This then requires that there is nonzero torsion and 
that the geometry is Riemann-Cartan.
Our final result is therefore that
{\it a Higgs mechanism for the spin connection is possible, 
but only in a Riemann-Cartan geometry.}
Constructing a ghost-free model with a propagating spin connection
is known to be a challenging problem.\cite{sc}
Evidently, incorporating Lorentz violation leads to the appearance of 
additional mass terms,
which in turn could create new possibilities for model building.

\section{Summary and Conclusions}

In theories with spontaneous Lorentz violation, 
up to ten NG modes can appear. 
They can all be incorporated naturally in the vierbein.  
For the vector bumblebee model, 
the Lorentz NG modes propagate like photons in an axial gauge. 
In principle, two Higgs mechanisms can occur,
one associated with broken diffeomorphisms,
the other with Lorentz symmetry.
While a usual Higgs mechanism (for diffeomorphisms)
does not occur involving the metric field,
an alternative Higgs mechanism can lead to the appearance of quadratic
metric terms in the Lagrangian.
If in addition the geometry is Riemann-Cartan, 
then a Higgs mechanism (for the Lorentz symmetry)
can occur in which the spin connection acquires a mass.
Clearly,
there are numerous phenomenological questions that arise in
these processes,  
all of which can all be pursued comprehensively using the SME.

\section*{Acknowledgments}

This work was supported  
by NSF grant PHY-0554663.

\end{document}